\documentclass[reprint,superscriptaddress,prb,aps]{revtex4-1}

\usepackage{amsmath}
\usepackage{amssymb}
\usepackage{graphicx}
\usepackage{bm}
\usepackage[colorlinks,linkcolor=blue, urlcolor=blue,citecolor=blue,bookmarks=false]{hyperref}

\begin{document}

\title{Mottness versus unit-cell doubling \protect\\ as the driver of the insulating state in 1\textit{T}-TaS$_{2}$}

\author{C. J. Butler}
\email{christopher.butler@riken.jp}
\affiliation{RIKEN Center for Emergent Matter Science, 2-1 Hirosawa, Wako, Saitama 351-0198, Japan}

\author{M. Yoshida}
\affiliation{RIKEN Center for Emergent Matter Science, 2-1 Hirosawa, Wako, Saitama 351-0198, Japan}

\author{T. Hanaguri}
\email{hanaguri@riken.jp}
\affiliation{RIKEN Center for Emergent Matter Science, 2-1 Hirosawa, Wako, Saitama 351-0198, Japan}

\author{Y. Iwasa}
\affiliation{RIKEN Center for Emergent Matter Science, 2-1 Hirosawa, Wako, Saitama 351-0198, Japan}
\affiliation{Quantum-Phase Electronics Center and Department of Applied Physics, The University of Tokyo, 7-3-1 Hongo, Bunkyo-ku, Tokyo 113-8656, Japan}

\maketitle
\textbf{
If a material with an odd number of electrons per unit cell turns out to be insulating, Mott localisation may be invoked as an explanation \cite{Mott1937,Imada1998}. This is widely accepted for the layered compound 1\textit{T}-TaS$_{2}$, which has a low-temperature insulating phase comprising charge order clusters with 13 unpaired orbitals each \cite{Wilson1975,Fazekas1979,Fazekas1980}. But if the stacking of layers doubles up the unit cell to include an even number of orbitals, the nature of the insulating state is ambiguous \cite{Ritschel2018,Lee2019}. Here, scanning tunnelling microscopy (STM) reveals two distinct terminations of the charge order in 1\textit{T}-TaS$_{2}$, the sign of such a double-layer stacking pattern. However, spectroscopy at both terminations allows us to disentangle unit-cell doubling effects and determine that Mott localisation alone is enough to drive gap formation. We also observe the collapse of Mottness at an extrinsically restacked termination, demonstrating that the microscopic mechanism of insulator-metal transitions \cite{Stojchevska2014,Hollander2015,Vaskivskyi2015,Yoshida2015,Vaskivskyi2016,Cho2016,Ma2016} lies in degrees of freedom of interlayer stacking.
}
\\

The undistorted high-temperature atomic structure of 1\textit{T}-TaS$_{2}$ is shown in Fig. 1a. Below $\sim$350 K, the Ta lattice within each layer undergoes a periodic in-plane distortion in which clusters of 13 Ta ions contract towards the central ion of the cluster forming a `Star-of-David' (SD) motif \cite{Wilson1975}. Upon cooling below $\sim$180 K, this pattern locks in to become commensurate with the atomic lattice and long range order emerges, described as a triangular $\sqrt{13}\times\sqrt{13}$ \textit{R}13.9$^\circ$ charge density wave (CDW) pattern, depicted in Fig. 1b. Within each of the SD clusters, 12 of the Ta 5\textit{d} orbitals at the periphery form 6 filled bands and leave a CDW gap \cite{Qiao2017}, stabilizing the distortion. The remaining orbital, according to band theory, should form a half-filled band, and the experimentally observed insulating behaviour is usually attributed to its localisation at the SD centre by strong electron-electron (\textit{e-e}) interactions \cite{Fazekas1979,Fazekas1980}. From this foundation it has been suggested that, since a Mott state in 1\textit{T}-TaS$_{2}$ realises a triangular lattice of localised $S = {}^1{\mskip -5mu/\mskip -3mu}_2$ spins, it might host a quantum spin liquid (QSL), an unusual phase of quantum electronic matter in which, due to geometric frustration and quantum fluctuations, the spins refuse to magnetically order even down to $T$ = 0 K  \cite{Balents2010,Law2017,Klanjsek2017,Ribak2017}.

The Mott state thought to exist in 1\textit{T}-TaS$_{2}$ is different from ordinary Mott insulators such as NiO, in that electrons localise not at the sites of the atomic crystal, but at the sites of the \textit{electronic} crystal, the lattice of SD clusters, and so it is called a `cluster Mott insulator'. As the SD clusters must be centred on Ta sites, the three-dimensional (3D) structure formed from the layering of 2D charge order lattices can be described with stacking vectors $\textbf{T}$ composed of the underlying Ta lattice vectors. Neglecting the S atomic layers sandwiching the Ta layer, there are only three symmetrically inequivalent stacking vectors: $\textbf{T}_{A}$ = \textbf{c}, $\textbf{T}_{B}$ = \textbf{a} + \textbf{c}, and $\textbf{T}_{C}$ = \textbf{a} + \textbf{b} + \textbf{c} (with the latter two each having a group of symmetrical equivalents). The impact of this stacking degree of freedom on the electronic structure of 1\textit{T}-TaS$_{2}$ was largely neglected until Ritschel \textit{et al} predicted, using \textit{ab initio} calculations, that different inter-layer stacking patterns could result in a metallic phase (for $\textbf{T}_{C}$ stacking) as an alternative to the well-known insulating phase (previously assumed to have $\textbf{T}_{A}$ stacking) \cite{Ritschel2015}. Going further, Ritschel \textit{et al} and Lee \textit{et al} recently challenged the rationale by which 1\textit{T}-TaS$_{2}$ was thought to be a Mott insulator, showing that if the stacking alternates between vectors $\textbf{T}_{A}$ and $\textbf{T}_{C}$ as previously suggested \cite{Tanda1984,Naito1984,Naito1986}, such that the new supercell includes two SD clusters, \textit{ab initio} calculations predict an insulator without the need to invoke strong \textit{e-e} interactions \cite{Ritschel2018,Lee2019}. More simply, if the electronic unit cell contains two SDs, the total number of electrons per cell is even, leaving the highest occupied band filled and allowing an insulator without invoking Mott. This introduces serious complication into the understanding of the insulating state in 1\textit{T}-TaS$_{2}$, and potentially undermines the foundations on which recent suggestions of a QSL state are built \cite{Law2017}.

\begin{figure*}
\centering
\includegraphics[scale=1]{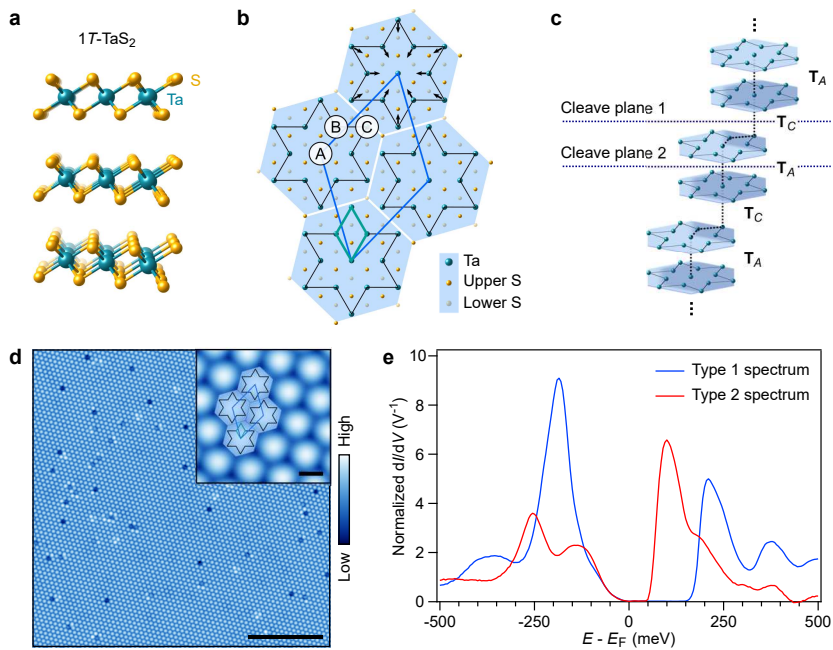}
\caption{\label{fig:1} \textbf{Overview of charge order, inter-layer stacking and cleaved surfaces in 1\textit{T}-TaS$_{2}$.} \textbf{a} The quasi-2D undistorted structure of 1\textit{T}-TaS$_{2}$. \textbf{b} The supercell describing the periodic SD distortion within a single 1\textit{T}-TaS$_{2}$ layer. The teal and blue rhombuses mark the 2D projections of the undistorted atomic unit cell, and the supercell after onset of the commensurate CDW, respectively. The labels \textit{A}, \textit{B}, and \textit{C} denote the possible sites atop which successive SD clusters can stack. \textbf{c} The SD stacking pattern currently discussed (S not shown), with two SDs per cell and two distinct cleavage planes, 1 \& 2. \textbf{d} Typical STM topography taken at a vacuum-cleaved 1\textit{T}-TaS$_{2}$ surface ($V$ = 250 mV, $I_{set}$ = 500 pA, scale bar 20 nm). The inset shows the correspondence between the topographic modulation and the SD cluster lattice (scale bar 1 nm). \textbf{e} Examples of conductance spectra of the two types observed at multiple cleaved surfaces. Typically, one type of spectrum or the other appears uniformly (except in the vicinity of defects) over $\sim$1 $\mu$m areas, unless a step-terrace morphology is observed. It will be shown below that the type 1 \& 2 spectra correspond to surfaces formed by cleavage at planes 1 \& 2 respectively.}
\end{figure*}

A consequence of the $\textbf{T}_{A}$, $\textbf{T}_{C}$, $\textbf{T}_{A}$, $\textbf{T}_{C}$\ldots (henceforth `\textit{ACAC}') stacking pattern is that there are two cleavage planes, as indicated in Fig. 1c, yielding two inequivalent surfaces amenable to investigation using STM. One plane is located between one $\textbf{T}_{A}$-stacked bilayer (BL) and another, and the other plane splits a single BL, leaving unpaired ($\textbf{T}_{C}$-stacked) layers. In this work, samples were cleaved, transferred to the STM and measured at temperatures far below the transition temperature at which the commensurate CDW sets in ($\sim$180 K), and the bulk structure of the CDW should be preserved such that measurements on a large number of cleaved surfaces may show evidence of the \textit{ACAC} pattern. Eight platelets of 1\textit{T}-TaS$_{2}$ were cleaved multiple times each, for a total of twenty-four investigated surfaces, the topographic image for one of which appears in Fig. 1d. Conductance spectra were acquired at defect-free locations on each sample. Spectra showing a gap in the density of states (DOS) of $\sim$150 meV, broadly consistent with those shown in previous STM reports \cite{Cho2016,Ma2016,Cho2017,Qiao2017} were observed on eighteen of the twenty-four surfaces (similar to the blue curve labelled `Type 1' in Fig. 1e). A different form of the DOS, with a smaller gap of $\sim$50 meV, was observed on the remaining six (`Type 2', the red curve in Fig. 1e). (We only consider the spectra acquired in the regions where the STM tip first arrived at the sample surface). We tentatively attribute the appearance of these two forms of DOS to the surfaces created by the two cleavage planes of the bulk stacking pattern. However, more information is needed to definitively assign each form of DOS to each cleaved surface, and we return to this below. If the number of cleavage planes of each type throughout the sample is roughly equal, as we assume, the deviation of the observed ratio from 1:1 may correspond to a difference in their associated surface formation energies, and the resulting rarity of the small-gap surface may explain its absence in previous reports. A small number of instances were observed where the two types of surface appeared side-by-side, for example on either side of a domain wall \cite{Cho2017} in the CDW pattern (see Supplementary Information).

\begin{figure*}
\centering
\includegraphics[scale=1]{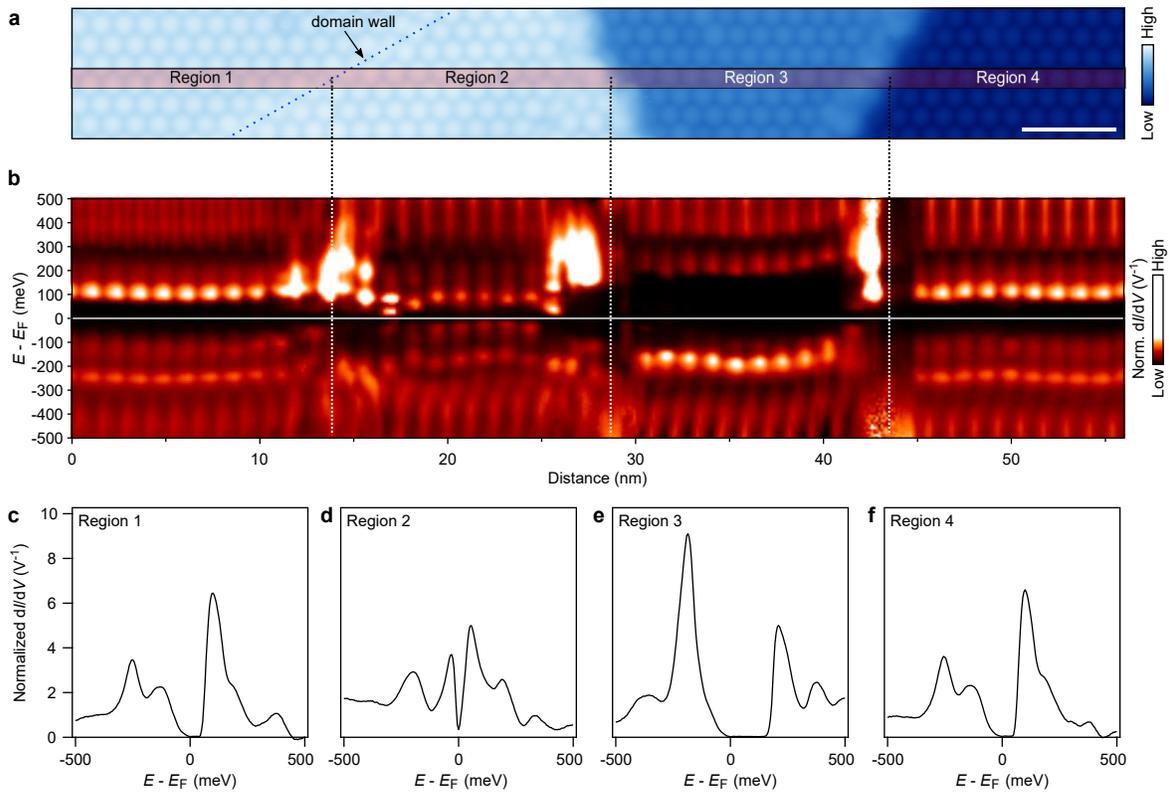}
\caption{\label{fig:2} \textbf{Conductance spectra across a step-terrace formation.} \textbf{a} A topographic image showing three terraces separated by two steps ($V$ = 250 mV, $I_{set}$ = 125 pA, scale bar 5 nm). The uppermost terrace at the left-hand-side is further divided into two regions separated by a domain wall, Regions 1 \& 2. \textbf{b} Spectroscopy along the path marked by the red-tinted rectangle in \textbf{a}, averaged over the rectangle's short axis. (The width of the rectangle was chosen so as to average over, approximately, the vertical projection of one CDW period.) \textbf{c}--\textbf{f} Representative spectra collected in each of the four regions, 1--4, shown in \textbf{a}. Leaving aside the metallic Region 2, the type of spectrum alternates layer-by-layer.}
\end{figure*}

Most revealingly, the two types were also observed side-by-side where single-layer steps allowed the simultaneous observation of multiple TaS$_{2}$ layers. Figure 2a shows a topographic image of three terraces, with the upper terrace featuring a domain wall (marked with a dark blue dotted line), so that four distinct regions are observed (labelled Regions 1--4). Tunnelling spectroscopy acquired along a path spanning the long axis of the topographic image (marked with a red-tinted rectangle) shows changes in the DOS spectrum upon each transition between regions (Fig. 2b). Representative spectra taken at a point within each of the four regions are shown in Fig. 2c--f. Region 2 shows a finite DOS at $E_{F}$, reminiscent of the `metallic mosaic’ phase which has been created locally using STM induced voltage pulses, \cite{Cho2016} with interlayer stacking effects suggested as a possible explanation \cite{Ma2016}. Briefly postponing the discussion of this metallic phase, we first note that the form of the DOS in the other three regions is seen to alternate from one terrace to the next, from a small gap (Region 1) to a large gap in the middle terrace (Region 3), and to the small gap again at the lowest terrace (Region 4). This alternating sequence is consistent with that expected for the \textit{ACAC} stacking shown in Fig. 1e. (Another, similar instance of the switching of electronic structure from one type to the other across a single layer step is shown in the Supplementary Information.)

\begin{figure*}
\centering
\includegraphics[scale=1]{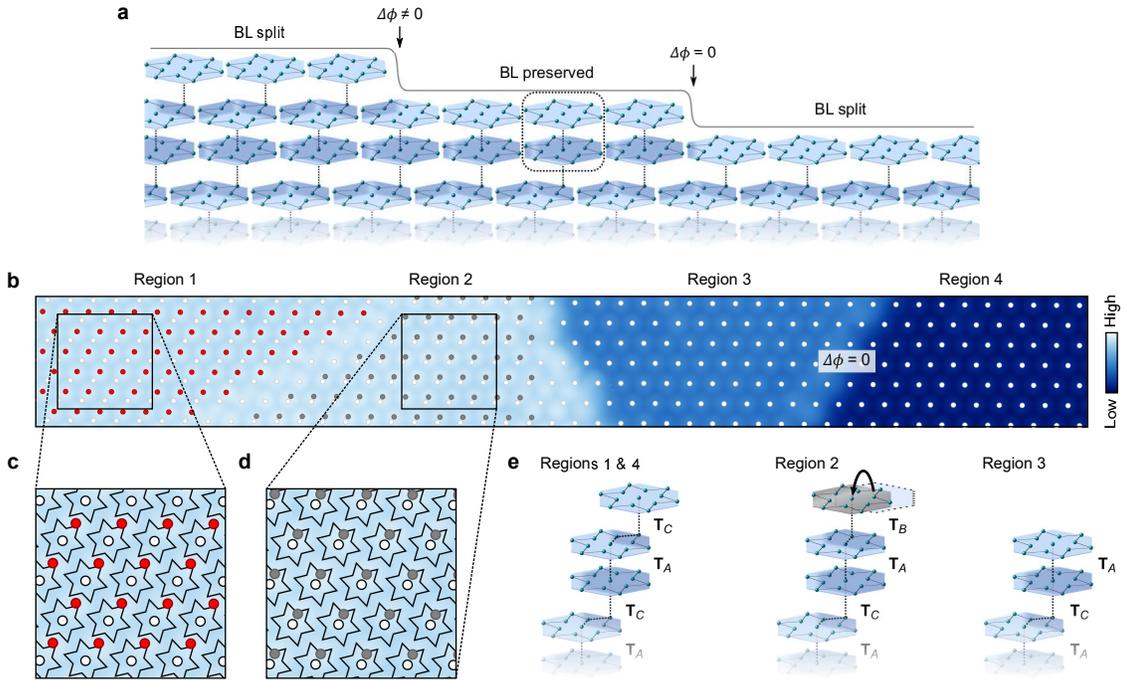}
\caption{\label{fig:3} \textbf{Identification of inter-layer stacking vectors.} \textbf{a} A schematic showing the in-plane displacements or phase shifts $\Delta \phi$, associated with steps of a pristine cleaved step-terrace morphology resulting from the \textit{ACAC} structure. $\Delta \phi$ alternates between zero and non-zero. \textbf{b} The same STM topography shown in Fig. 2, with the SD cluster centres in Regions 3 \& 4 marked with white dots. The lower two terraces are shown to be in phase with each other, indicating that they belong to the same BL. The lattice from the lower two terraces is extended to the left-hand-side of the field-of-view. The cluster centres in Regions 1 \& 2 are shown with red and grey dots respectively. \textbf{c} and \textbf{d} The relative positions of the top terrace clusters with respect to the SD motifs of the underlying BL. \textbf{e} Schematic depictions of the inter-layer stacking configuration for each Region, to which we attribute the DOS spectra shown in Fig. 2. The layer in Region 2 which we interpret as extrinsically re-stacked is shown in grey.}
\end{figure*}

With a view to establishing which type of surface corresponds to which of the cleavage planes in the \textit{ACAC} stacking pattern, we note that single-layer steps should result in an in-plane displacement, or phase jump $\Delta \phi$, of the 2D projected CDW pattern from one terrace to the next, which should alternate between zero and non-zero (specifically \textbf{a} + \textbf{b}, the in-plane projection of $\textbf{T}_{C}$), as is shown in Fig. 3a. The step between Regions 3 \& 4 realises the former case: In Fig. 3b, the SD centres in these two regions are highlighted with an array of white dots, showing clearly the absence of an in-plane displacement and hence indicating that in Region 3 the stacking pattern terminates with an intact BL at the surface, and that in Region 4, an unpaired layer of SD clusters remains.

The array of white dots is extended over to the left-hand-side of the image. The centres of the SD lattice in the two regions of the upper terrace are overlaid with this reference lattice to show the approximate in-plane shift. Additional atomically resolved topographic imaging in Region 1 (see Supplementary Information) is used to determine the orientation of the SD clusters, depicted in the zoom-in images of Figs. 3c,d. The in-plane components of the stacking vectors for Regions 1 \& 2 are then discernible. In Region 1 the stacking vector is shown to correspond closely to $\textbf{T}_{C}$, as expected for the \textit{ACAC} stacking pattern, with a discrepancy well below one atomic lattice constant. The stacking vector between the metallic region, Region 2, and the underlying BL is also determined, and corresponds closely to the $\textbf{T}_{B}$. As this type of metallic region was observed nowhere else throughout measurements on twenty-three other samples, we posit that it is an outcome of an extrinsic process during formation of the step-edge, such as a brief lifting of the uppermost TaS$_{2}$ layer and re-stacking of its CDW into a metastable configuration.

Taken together, the observations presented in Figs. 2 and 3 form a picture which is indeed consistent with the previously suggested \textit{ACAC} stacking pattern, and allow us to establish a link between the surface CDW configuration and the surface local DOS, as summarized in Fig. 3e. The spectrum observed in Region 3 with the relatively large gap of $\sim$150 meV has been reported in multiple STM works. We now suggest that this spectrum signifies a surface where the CDW terminates with the $\textbf{T}_{A}$-stacked BL intact, and can be thought of as the bulk-like termination of the CDW (\textit{i.e.} without major structural or electronic reconfiguration with the formation of the surface). The fact that this surface was the most common outcome from cleavage (18 out of 24) indicates the energetic favourability of cleaving between BLs, rather than through them, suggesting non-negligible intra-BL bonding.

The surface of unpaired SD clusters (Regions 1 \& 4) represents a new and perhaps qualitatively distinct system. It satisfies the conditions under which 1\textit{T}-TaS$_{2}$ was originally thought to be a Mott insulator: A system with an odd number of electrons per (surface) unit cell, and yet which is insulating. This re-affirms the importance of strong \textit{e-e} interactions in the insulating state of 1\textit{T}-TaS$_{2}$, complementing the recent observation of doublon excitations characteristic of a Mott state \cite{Ligges2018}. An insulating state in the unpaired layer also leaves some room for the persistence, at surfaces and possibly bulk stacking faults, of a QSL ground state despite possible inter-layer singlet formation within the bulk BLs \cite{Law2017}. It also suggests that the metallic state also observed (Region 2 in Fig. 2) is properly described as a `Mottness-collapsed' state, in which the Mottness ratio $U/t$ has been reduced either by increased screening of the on-site Coulomb repulsion $U$, or due to an increase of the inter-cluster overlap integral $t$. It is noteworthy that although the DOS at $E_{F}$ is non-zero, the residual pseudogap-like features around $E_{F}$ suggest the persistence of strong \textit{e-e} interactions in this metallic state.

Questions arise about the detailed mechanisms in play at these distinct surfaces -- Why does the unpaired cluster layer have a smaller gap than the paired layer, and why does Mottness break down leading to metallicity for $\textbf{T}_{B}$ stacking? Further investigations of the detailed behaviour of the 3D electronic correlations in 1\textit{T}-TaS$_{2}$, especially at its surface terminations and also in few-layer or monolayer form, may prove fruitful in understanding the nature of its insulating state and its potentially useful metal-insulator transitions.

\subsection*{Methods}
Crystals of 1\textit{T}-TaS$_{2}$ were synthesized using a chemical vapor transport method described previously \cite{Tani1981}, with 2 \% excess S. Samples were cleaved in ultra-high vacuum ($\sim$10$^{-10}$ Torr) at 77 K and quickly inserted into a Unisoku 1300 low temperature STM system, of which the STM head has been replaced with a homemade one \cite{Hanaguri2006}. All measurements were performed at a temperature of 1.5 K. STM tips were formed using electro-chemical etching of tungsten wire, and after insertion into UHV, were cleaned and characterized using field ion microscopy followed by careful conditioning on a clean Cu(111) surface. STM topography images were collected in constant-current mode. For conductance curves and spectroscopic mapping on 1\textit{T}-TaS$_{2}$ surfaces, the lock-in technique with a bias modulation of amplitude $V_{mod}$ = 10 mV and frequency $f_{mod}$ = 617.3 Hz was used.

The $dI/dV$ spectroscopy data in Figs. 1 \& 2, including the spatially resolved data shown in Fig. 2b, were normalized according to the current signal at $V$ = 500 mV. This somewhat compensates for the large difference in raw signal intensity caused by differing tip-sample distances while scanning at $V$ = 250 mV over areas with large or small spectral gaps.

\subsection*{Acknowledgements}
We are grateful to Y. Kohsaka, T. Machida and P. A. Lee for helpful discussions. C.J.B. gratefully acknowledges support from RIKEN's SPDR fellowship. This work was supported in part by JSPS KAKENHI grant numbers JP18K13511, JP19H00653 and JP19H01855.

\subsection*{Author contributions}
T.H. and Y.I. conceived the project, and M.Y. and Y.I. synthesized the 1\textit{T}-TaS$_{2}$ crystals. C.J.B. performed the STM measurements with assistance from T.H., and prepared the manuscript with input from all authors.


\begin{thebibliography}{99}

\bibitem{Mott1937}
Mott, N. F. \& Peierls, R.
Discussion of the paper by de Boer and Verway.
\textit{Proc. Phys. Soc. Lond.} \textbf{49,} 72 (1937).
\url{https://doi.org/10.1088/0959-5309/49/4S/308}


\bibitem{Imada1998}
Imada, M., Fujimori, A. \& Tokura, Y.
Metal-insulator transitions.
\textit{Rev. Mod. Phys.} \textbf{70,} 1039 (1998).
\url{https://doi.org/10.1103/RevModPhys.70.1039}


\bibitem{Wilson1975}
Wilson, J. A., Di Salvo, F. J. \& Mahajan, S.
Charge-density waves and superlattices in the metallic layered transition metal dichalcogenides.
\textit{Adv. Phys.} \textbf{24,}  117--201 (1975).
\url{https://doi.org/10.1080/00018737500101391}


\bibitem{Fazekas1979}
Fazekas, P. \& Tosatti, E.
Electrical, structural and magnetic properties of pure and doped 1T-TaS$_{2}$.
\textit{Phil. Mag. B} \textbf{39,} 229--244 (1979).
\url{https://doi.org/10.1080/13642817908245359}


\bibitem{Fazekas1980}
Fazekas, P. \& Tosatti, E.
Charge carrier localization in pure and doped 1T-TaS$_{2}$.
\textit{Physica B \& C} \textbf{99,} 183--187 (1980).
\url{https://doi.org/10.1016/0378-4363(80)90229-6}


\bibitem{Ritschel2018}
Ritschel, T., Berger, H. \& Geck, J.
Stacking-driven gap formation in layered 1\textit{T}-TaS$_{2}$.
\textit{Phys. Rev. B} \textbf{98,} 195134 (2018).
\url{https://doi.org/10.1103/PhysRevB.98.195134}


\bibitem{Lee2019}
Lee, S.-H., Goh, J. S. \& Cho, D.
Origin of the Insulating Phase and First-Order Metal-Insulator Transition in 1\textit{T}-TaS$_{2}$.
\textit{Phys. Rev. Lett.} \textbf{122,} 106404 (2019).
\url{https://doi.org/10.1103/PhysRevLett.122.106404}


\bibitem{Stojchevska2014}
Stojchevska, L. \textit{et al.}
Ultrafast switching to a stable hidden quantum state in an electronic crystal.
\textit{Science} \textbf{344,} 177–180 (2014).
\url{https://doi.org/10.1126/science.1241591}


\bibitem{Hollander2015}
Hollander, M. J. \textit{et al.}
Electrically Driven Reversible Insulator-Metal Transition in 1T-TaS$_{2}$.
\textit{Nano Lett.} \textbf{15,} 1861--1866 (2015).
\url{https://doi.org/10.1021/nl504662b}


\bibitem{Vaskivskyi2015}
Vaskivskyi, I. \textit{et al.}
Controlling the metal-to-insulator relaxation of the metastable hidden quantum state in 1T-TaS$_{2}$.
\textit{Sci. Adv.} \textbf{1,} e1500168 (2015).
\url{https://doi.org/10.1126/sciadv.1500168}


\bibitem{Yoshida2015}
Yoshida, M., Suzuki, R., Zhang, Y., Nakano. M. \& Iwasa, Y.
Memristive phase switching in two-dimensional 1T-TaS$_{2}$.
\textit{Sci. Adv.} \textbf{1,} e1500606 (2015).
\url{https://doi.org/10.1126/sciadv.1500606}


\bibitem{Vaskivskyi2016}
Vaskivskyi, I. \textit{et al.}
Fast electronic resistance switching involving hidden charge density wave states.
\textit{Nat. Communs.} \textbf{7,} 11442 (2016).
\url{https://doi.org/10.1038/ncomms11442}


\bibitem{Cho2016}
Cho, D. \textit{et al.}
Nanoscale manipulation of the Mott insulating state coupled to charge order in 1\textit{T}-TaS$_{2}$.
\textit{Nat. Communs.} \textbf{7,} 10453 (2016).
\url{https://doi.org/10.1038/ncomms10453}


\bibitem{Ma2016}
Ma, L. \textit{et al.}
A metallic mosaic phase and the origin of Mott-insulating state in 1T-TaS$_{2}$.
\textit{Nat. Communs.} \textbf{7,} 10956 (2016).
\url{https://doi.org/10.1038/ncomms10956}


\bibitem{Balents2010}
Balents, L.
Spin liquids in frustrated magnets.
\textit{Nature} \textbf{464,} 199--208 (2010).
\url{https://doi.org/10.1038/nature08917}


\bibitem{Law2017}
Law, K. T. \& Lee, P. A.
1T-TaS$_{2}$ as a quantum spin liquid. \textit{Proc. Natl. Acad. Sci.}
\textbf{114} (27), 6996--7000 (2017).
\url{https://doi.org/10.1073/pnas.1706769114}


\bibitem{Klanjsek2017}
Klanj\v{s}ek, M. \textit{et al.}
A high-temperature quantum spin liquid with polaron spins.
\textit{Nat. Phys.} \textbf{13,} 1130--1134 (2017).
\url{https://doi.org/10.1038/nphys4212}


\bibitem{Ribak2017}
Ribak, A. \textit{et al.}
Gapless excitations in the ground state of 1\textit{T}-TaS$_{2}$.
\textit{Phys. Rev. B} \textbf{96,} 195131 (2017).
\url{https://doi.org/10.1103/PhysRevB.96.195131}


\bibitem{Ritschel2015}
Ritschel, T. \textit{et al.}
Orbital textures and charge density waves in transition metal dichalcogenides.
\textit{Nat. Phys.} \textbf{11,} 328--331 (2015).
\url{https://doi.org/10.1038/nphys3267}


\bibitem{Tanda1984}
Tanda, S., Sambongi, T., Tani, T. \& Tanaka, S.
X-Ray Study of Charge Density Wave Structure of 1T-TaS$_{2}$.
\textit{J. Phys. Soc. Jpn.} \textbf{53,} 476--479 (1984).
\url{https://doi.org/10.1143/JPSJ.53.476}


\bibitem{Naito1984}
Naito, M., Nishihara, H. \& Tanaka, S.
Nuclear quadrupole resonance in the charge density wave state of 1\textit{T}-TaS$_{2}$.
\textit{J. Phys. Soc. Jpn.} \textbf{53,} 1610--1613 (1984).
\url{https://doi.org/10.1143/JPSJ.53.1610}


\bibitem{Naito1986}
Naito, M., Nishihara, H. \& Tanaka, S.
Nuclear magnetic resonance and nuclear quadrupole resonance study of $^{181}$Ta in the commensurate charge density wave state of 1\textit{T}-TaS$_{2}$.
\textit{J. Phys. Soc. Jpn.} \textbf{55,} 2410--2421 (1986).
\url{https://doi.org/10.1143/JPSJ.55.2410}


\bibitem{Cho2017}
Cho, D. \textit{et al.}
Correlated electronic states at domain walls of a Mott-charge-density-wave insulator 1\textit{T}-TaS$_{2}$.
\textit{Nat. Communs.} \textbf{8,} 392 (2017).
\url{https://doi.org/10.1038/s41467-017-00438-2}


\bibitem{Qiao2017}
Qiao, S. \textit{et al.}
Mottness Collapse in 1T-TaS$_{2-x}$Se$_{x}$ Transition-Metal Dichalcogenide: An Interplay between Localized and Itinerant Orbitals.
\textit{Phys. Rev. X} \textbf{7,} 041054 (2017).
\url{https://doi.org/10.1103/PhysRevX.7.041054}


\bibitem{Ligges2018} Ligges, M. \textit{et al.}
Ultrafast Doublon Dynamics in photoexcited 1\textit{T}-TaS$_{2}$.
\textit{Phys. Rev. Lett.} \textbf{120,} 166401 (2018).
\url{https://doi.org/10.1103/PhysRevLett.120.166401}


\bibitem{Tani1981}
Tani, T., Okajima, K., Itoh, T. \& Tanaka S.
Electronic transport properties in 1T-TaS$_{2}$.
\textit{Physica B \& C} \textbf{105,} 127--131 (1981).
\url{https://doi.org/10.1016/0378-4363(81)90230-8}


\bibitem{Hanaguri2006}
Hanaguri, T.
Development of high-field STM and its application to the study on magnetically tuned criticality in Sr$_{3}$Ru$_{2}$O$_{7}$.
\textit{J. Phys. Conf. Ser.} \textbf{51,} 514 (2006).
\url{https://doi.org/10.1088/1742-6596/51/1/117}


\end{thebibliography}
\end{document}


\title{Supplementary Information for \protect\\ Mottness versus unit-cell doubling \protect\\ as the driver of the insulating state in 1\textit{T}-TaS$_{2}$}

\author{C. J. Butler}
\email{christopher.butler@riken.jp}
\affiliation{RIKEN Center for Emergent Matter Science, 2-1 Hirosawa, Wako, Saitama 351-0198, Japan}

\author{M. Yoshida}
\affiliation{RIKEN Center for Emergent Matter Science, 2-1 Hirosawa, Wako, Saitama 351-0198, Japan}

\author{T. Hanaguri}
\email{hanaguri@riken.jp}
\affiliation{RIKEN Center for Emergent Matter Science, 2-1 Hirosawa, Wako, Saitama 351-0198, Japan}

\author{Y. Iwasa}
\affiliation{RIKEN Center for Emergent Matter Science, 2-1 Hirosawa, Wako, Saitama 351-0198, Japan}
\affiliation{Quantum-Phase Electronics Center and Department of Applied Physics, The University of Tokyo, 7-3-1 Hongo, Bunkyo-ku, Tokyo 113-8656, Japan}

\maketitle


\section{Supplementary discussion}

\subsection{I. Additional side-by-side observations of distinct terminations.} 

In this work, 24 surfaces were investigated and the density-of-states (DOS) spectrum with the larger gap ($\sim$150 meV), and with the smaller gap ($\sim$50 meV) were observed 18 and 6 times, respectively, considering only the spectrum found at the landing zone of the tip in each case. At 3 surfaces the type of spectrum was seen to change upon crossing some form of boundary after moving away from the landing zone. One example, a pair of single-layer steps, is described in the main text (Figs. 2 \& 3). The others, a CDW domain wall (DW), and another single-layer step, are described here.

Domain walls -- in-plane phase discontinuities in the CDW -- were often found at pristine cleaved surfaces. In principle, such DWs can offer opportunities to investigate the effects of inter-layer stacking: If one CDW layer with a DW is overlaid atop another layer without one, the phase jump \textit{must} result in different inter-layer stacking on either side, and be accompanied by a change in the DOS if the stacking vectors on each side are symmetrically inequivalent (e.g. $\textbf{T}_{A}$ versus $\textbf{T}_{C}$). In practice however, the same type of spectrum (either large or small gap) was almost always observed on both sides of DWs. This may be because the DWs extend downwards below the surface such that the surface layer and buried layers all have matching phase jumps, or because the change in stacking from one side to another is merely from one among the group of symmetrically equivalent $\textbf{T}_{C}$ vectors to another. However, on one occasion, shown in Fig. S1 below, the two types of spectrum were observable in the same field of view, in different domains separated by an unusual DW.

\begin{figure}[!hb]
\centering
\includegraphics[scale=1]{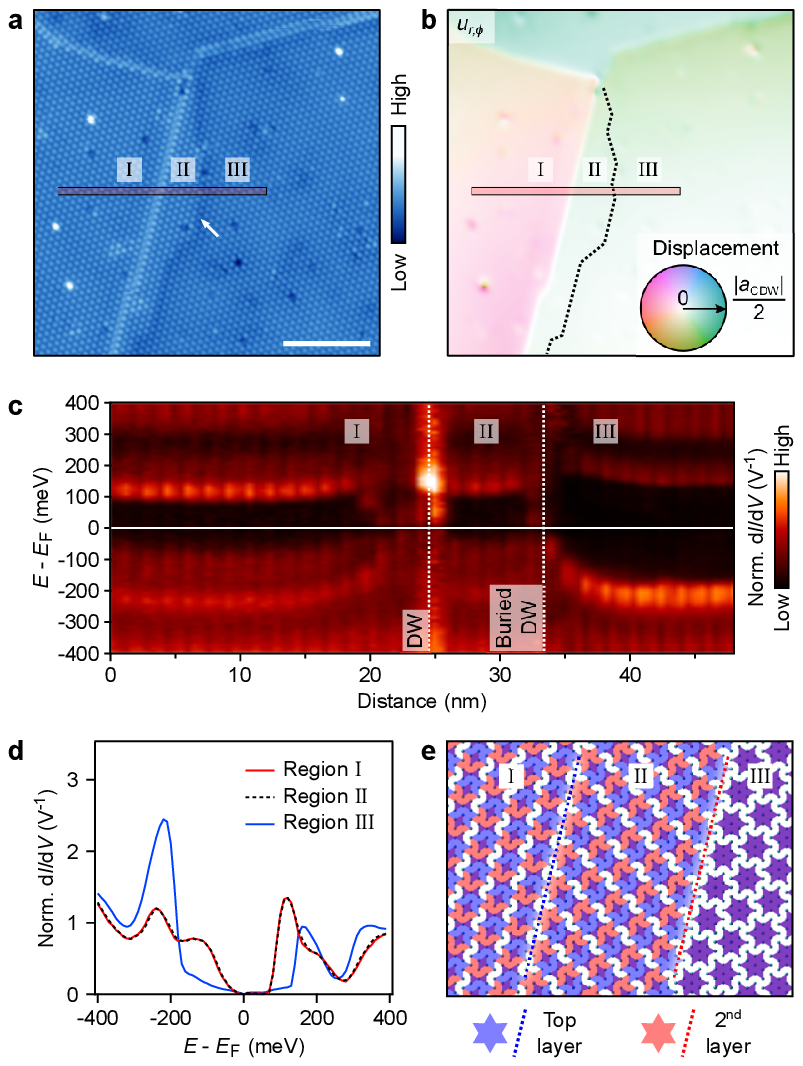}
\caption{\label{fig:1} \textbf{Change in electronic structure with change of stacking across a buried DW.} (\textbf{a}) A topographic image showing multiple DWs found at a pristine cleaved surface ($V$ = 500 mV, $I_{set}$ = 10 pA, scale bar 20 nm). (\textbf{b}) The displacement map $\textit{u}_{r,\phi}(\textbf{x,y})$ (with displacement taken relative to a point in the bottom-right corner) highlighting the in-plane phase jumps of the CDW pattern of the surface layer at the bright DWs. At the position of the buried DW, marked by the black dotted line, there is no displacement. (\textbf{c}) Spatially resolved tunnelling spectroscopy taken along the long axis of the red tinted rectangle in (a) \& (b), crossing both top layer DW and the buried one. A change in DOS is observed only at the buried DW. (\textbf{d}) Representative tunnelling spectra acquired in Regions I--III. (\textbf{e}) A schematic cartoon of the possible stacking configurations in each of the three regions.}
\end{figure}

In Figure S1(a), ordinary DWs appear as bright linear features, along with another feature appearing as a dark line or depression (indicated by the white arrow). Here we use a partial implementation of the Lawler-Fujita algorithm \cite{Lawler2010} in order to visualize the in-plane phase discontinuities in the surface CDW pattern. Essentially the lock-in phase detection technique is applied to the image, yielding a map of the spatially varying phase discrepancy (displacement) between the observed CDW modulation and an ideal reference modulation. Figure S1(b) shows the resulting displacement field $\textit{u}_{r,\phi}(\textbf{x,y})$. (The displacement is taken with reference to the bottom-right corner of the image, and expressed in polar coordinates.) The sharp changes in hue signify in-plane phase jumps. Importantly, although phase jumps are clear for the bright DWs, none is seen for the dark meandering feature (marked by a black dotted line). We therefore interpret this dark feature as a \textit{buried} DW, with a continuous-phase layer of the CDW overlaid on top. Such a feature must be accompanied by a change of the inter-layer stacking.

\begin{figure}[!ht]
\centering
\includegraphics[scale=1]{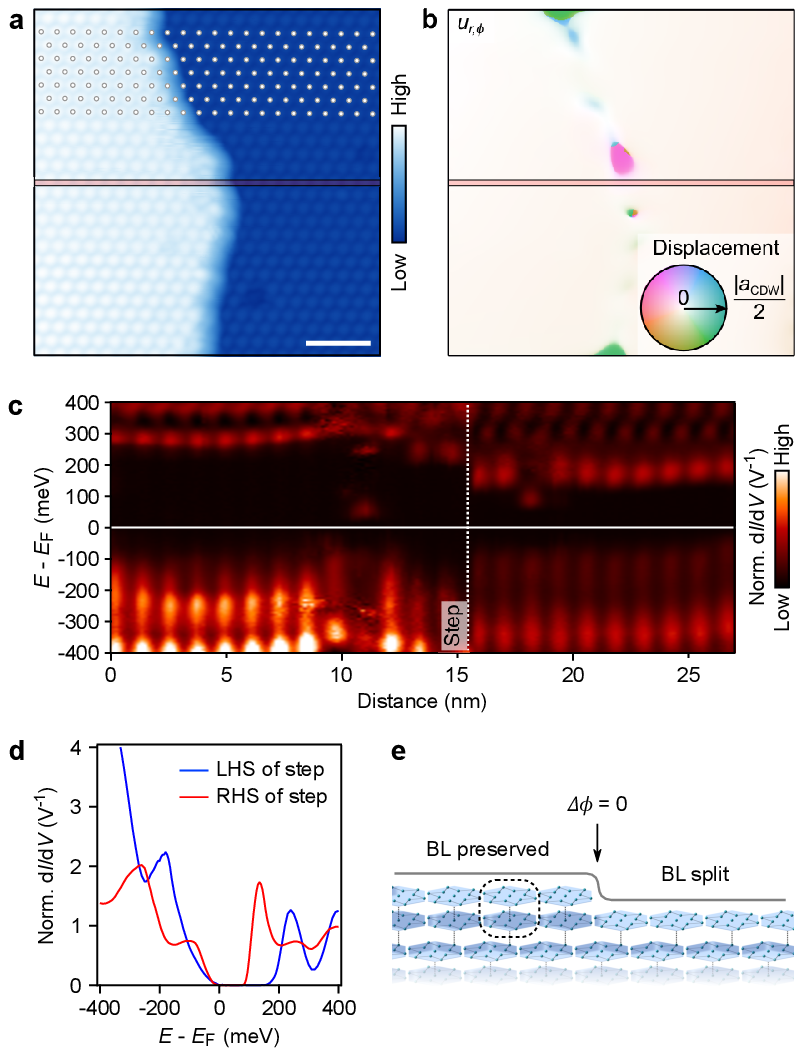}
\caption{\label{fig:2} \textbf{Partial replication of step-terrace measurements in Figs 2 \& 3 accompanying the main text.} (\textbf{a}) A topographic image showing two terraces separated by a single-layer step ($V$ = -500 mV, $I_{set}$ = 500 pA, scale bar 5 nm). (\textbf{b}) Map of the displacement field $\textit{u}_{r,\phi}(\textbf{x,y})$ (with displacement taken with respect to a reference point in the top-left corner), showing an absence of in-plane phase jump (displacement) from the upper to the lower terrace. (\textbf{c}) Tunneling spectroscopy along the path marked by the red-tinted rectangle in (a), averaged over the rectangle's short axis, and (\textbf{d}) typical spectra acquired on each terrace. (\textbf{e}) Schematic showing the correspondence between the above observations and the \textit{ACAC} stacking pattern.}
\end{figure}

Figure S1(c) shows the spatially resolved tunnelling spectroscopy collected along the red-tinted rectangle shown in (a) \& (b). The relatively small gap is seen on the left-hand-side (Region I), and does not change when crossing the ordinary DW into Region II. The spectrum does change when crossing the \textit{buried} DW in to Region III however. This provides strong evidence explicitly linking the change of DOS to a change of the inter-layer stacking vector \textbf{T}, although the exact stacking vectors cannot be inferred only from this measurement. Representative spectra from each of the Regions I--III are shown in Fig. S1(d).

A possible scenario which could realise the above observations in shown in Figure S1(e). The top and second layer CDW patterns are depicted with blue and red star motifs, respectively. The DW in the top layer (blue dashes) changes the inter-layer stacking from one of the group of symmetrically equivalent $\textbf{T}_{C}$ vectors to another (from Region I to II). The DW in the lower layer (red dashes) changes the stacking from $\textbf{T}_{C}$ to $\textbf{T}_{A}$ (from Region II to III), which would induce the observed change in DOS from the small gap (unpaired SD layer) to the large gap (intact BL).

Figure S2 shows a partial replication of the results in Figs 2 \& 3 accompanying the main text, acquired at the cleaved surface of a different sample. This provides supporting evidence for the distinct DOS spectra of the two terminations of the BL stacking pattern. An array of white dots (shown in Fig. S2(a)), along with the maps of displacement field $\textit{u}_{r,\phi}(\textbf{x,y})$ (Fig. S2(b)), show an absence of a displacement in the in-plane projected CDW pattern at the step. Within the \textit{ACAC} stacking scheme, we interpret this as a step from the top of an intact BL (left-hand-side), down to the top of an unpaired layer (right-hand-side), formerly the bottom half of the same BL. The concurrent change in DOS from the larger to the smaller gap, shown in Fig S2(c) \& (d), is consistent with the findings described in the main text for Figs 2 \& 3. Figure S2(e) gives a schematic showing a possible surface configuration of SD layers for this case.

\subsection{II. Determination of SD cluster orientation shown in Figs. 3(c) \& (d).}

At the step-terrace formation discussed in the main text, the approximate in-plane displacements of Regions 1 \& 2 with respect to the underlying bilayer (BL) seen in Regions 3 \& 4, are observed. But in order to interpret these displacements in terms of stacking vectors \textbf{T}, the relative angular orientations of the atomic and CDW lattices must be determined.

The $\sqrt{13}\times\sqrt{13}$ \textit{R}13.9$^\circ$ superstructure of Star-of-David (SD) clusters is chiral, and can manifest in two configurations, shown in Fig. S3(a) \& (b), related to each other by a mirror symmetry operation. They can be described with angles $\pm13.9^\circ$ between the atomic and CDW lattice vectors.

\begin{figure*}
\centering
\includegraphics[scale=1]{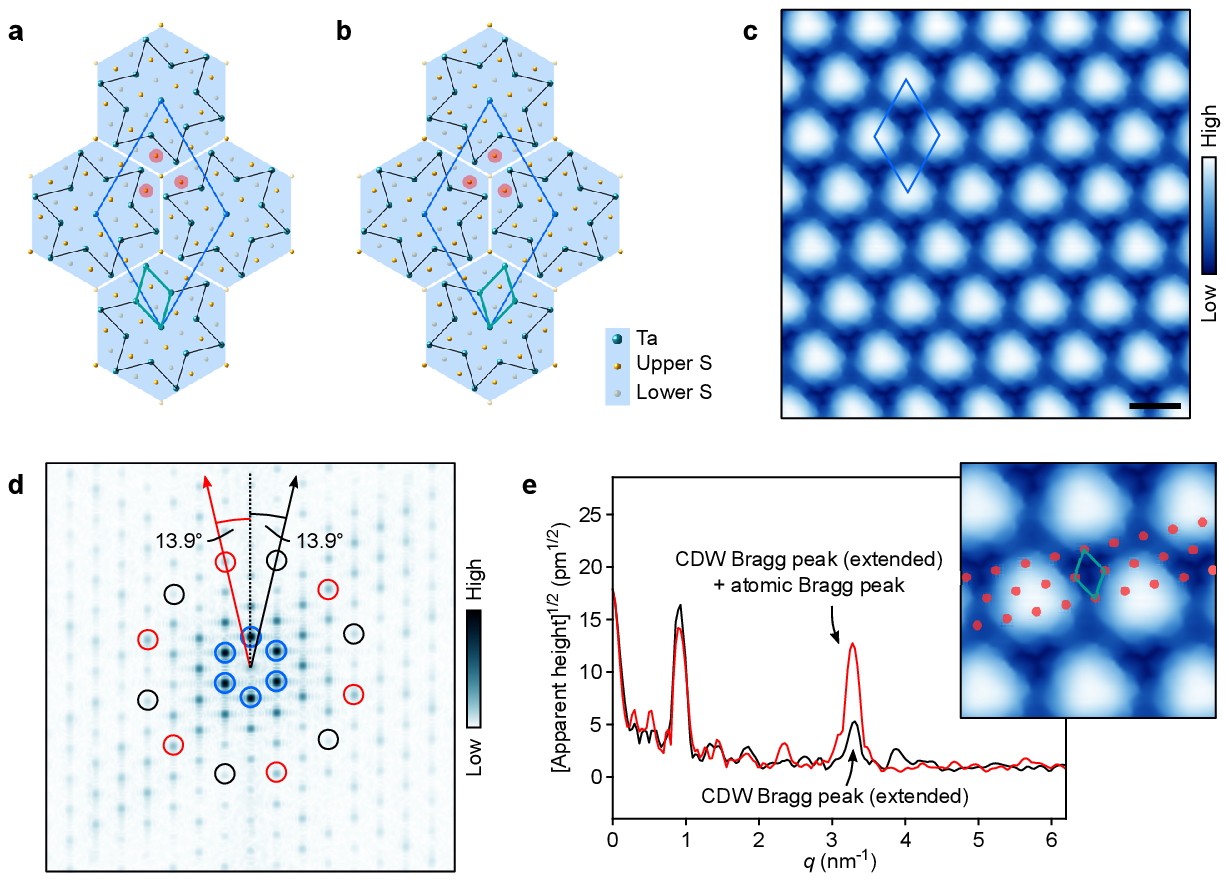}
\caption{\label{fig:3} \textbf{Determination of relative orientations of atomic and CDW lattices.} (\textbf{a}) and (\textbf{b}) Depictions of two possible orientations of the underlying atomic lattice with respect to the CDW lattice. (\textbf{c}) Atomically resolved STM topography acquired in Region 1 of Figs. 2 \& 3 discussed in the main text. (\textbf{d}) The 2D Fourier transform of the topography. First order reciprocal lattice peaks for the CDW are circled in blue. The two possible sets of locations for the atomic Bragg peaks are circled in black and red. (\textbf{e}) Line cuts through the Fourier transform at $\pm13.9^\circ$ with respect to the axis of the first order CDW reciprocal lattice peaks. The inset depicts the likely S surface lattice positions as red dots, for comparison with panel (a).}
\end{figure*}

Atomically resolved topography acquired in the same domain as Region 1 of Figs. 2 \& 3 accompanying the main text, shown in Fig. S3(c), became obtainable after the STM tip was contaminated by unstable debris. In the corresponding 2D Fourier transform image (Fig. S3(d)), the first order peaks of the CDW reciprocal lattice are circled in blue. Higher order peaks of the extended CDW reciprocal lattice appear over the entire \textbf{q}-space field of view. The atomic Bragg peaks are expected to lie on top of one set of these higher order CDW peaks, at the positions marked by sets of black or red circles in Fig. S3(d). Taking linecuts at $\pm13.9^\circ$ rotation away from the direction of a first order CDW reciprocal lattice peak (Fig. S3(e)), we find that the intensity at the expected position of the atomic Bragg peak is higher at $+13.9^\circ$ than at $-13.9^\circ$. The corresponding real-space atomic lattice is depicted in the inset, with red dots. In the 1\textit{T}-TaS$_{2}$ structure the same relative orientation between the CDW and atomic lattices can be assumed for each of the three observed terraces. At this point the in-plane in-plane displacements discussed in the main text can be interpreted in terms of the atomic lattice sites within the SD clusters depicted in panel (a), and the stacking vectors \textbf{T} can be determined.